\documentclass[twocolumn,preprintnumbers,floatfix,prb,superscriptaddress]{revtex4}
\usepackage{graphicx}
\usepackage{graphicx}% Include figure files
\usepackage{dcolumn} % Align table columns on decimal point
\usepackage{bm}
\usepackage{epsfig}
\begin{document} 

\title{Metal-insulator transition through a semi-Dirac point \\
in oxide nanostructures: VO$_2$ (001) layers confined within TiO$_2$}

\author{Victor Pardo}
\email{victor.pardo@usc.es}
\affiliation{Department of Physics,
  University of California, Davis, CA 95616
}
\affiliation{
Departamento de F\'{\i}sica Aplicada, Universidade
de Santiago de Compostela, E-15782 Santiago de Compostela,
Spain
}

\author{Warren E. Pickett}
\email{wepickett@ucdavis.edu}
\affiliation{Department of Physics,
  University of California, Davis, CA 95616
}

\begin{abstract}

Multilayer (TiO$_2$)$_m$/(VO$_2$)$_n$ nanostructures ($d^1$ - $d^0$ interfaces with no polar
discontinuity) show a metal-insulator transition with respect to the VO$_2$ layer thickness
in first principles calculations. 
For $n$ $\geq$ 5 layers, the system becomes metallic, while being insulating for $n$ = 1 and 2. 
The metal-insulator transition occurs through a semi-Dirac point phase for $n$ = 3 and 4, in which 
the Fermi surface is point-like and the electrons behave as massless along the zone diagonal in 
k-space and as massive fermions along the perpendicular direction. We provide an analysis of
the evolution of the electronic structure through this unprecedented insulator-to-metal
transition, and identify it as resulting from quantum confinement producing a non-intuitive 
orbital ordering on the V $d^1$ ions, rather than being a specific
oxide interface effect.  Spin-orbit coupling does not 
destroy the semi-Dirac point for the calculated ground state, where the spins are aligned along
the rutile $c$-axis, but it does open a substantial gap if the spins lie in the basal plane. 
% The ground state has 
% very small orbital moments, oriented along the 001 rutile direction and the changes in the 
% band structure introduced by spin-orbit coupling are very small.

\end{abstract}

\maketitle

\section{Background}

VO$_2$ in its bulk form displays a metal-insulator transition (MIT) which occurs near RT (at approximately 340 K).\cite{vo2_mit} This change in the electronic structure is accompanied by a structural distortion that involves the dimerization (to a low-temperature monoclinic insulating phase) of the V chains that form in the rutile structure, leading to a V-V singlet state accompanied by the expected reduction of the magnetic susceptibility.\cite{vo2_peierls1} The direct V-V exchange along the chains forms spin-singlets at low temperature, that break up above 340 K leading to a structural distortion, with dimers no longer occurring and V-V distances becoming uniform along the rutile c-axis. This MIT has been widely studied,  with effort given to tuning the temperature of the transition,\cite{if_apl} because its proximity to RT makes it usable for various applications.\cite{vo2_gas_sensor,vo2_nanobeams}

Recently oxide interfaces (IF) have shown interesting (and not yet well understood) MITs. One of these happens with a few layers of LaAlO$_3$ (LAO) on top of a SrTiO$_3$ (STO) substrate. Here, the effects of having a polar IF come into play, producing an electric field within the LAO layer that raises the energies (bands) of successive layers. Calculations show that the onset of the overlap of valence band O $2p$ states at the surface with Ti $3d$ conduction states on the substrate produces an insulator-to-metal transition in the heterostructure around 4 layers of LAO grown on an STO substrate,\cite{pentcheva2006,siemons,freeman,pentcheva2008,willmott} in agreement with experimental observations.

In this paper we will be dealing with a different mechanism for a MIT in an oxide nanostructure. This transition again happens at very small thicknesses, and involves several factors including the quantum confinement of the electronic states in the VO$_2$ layers. Experimentally, it was shown that the MIT seen in bulk VO$_2$ is inhibited in these films of VO$_2$ deposited on TiO$_2$. Below 5 nm,\cite{vo2_jap} there is not enough cooperativity along the z-axis to produce a dimerization of the V-V chains, thus not allowing for an insulating phase to develop. In addition, the lattice strain induced by the substrate might be strong enough in thin layers to prevent the MIT from occurring.

However, when the effects of quantum confinement come into play (for thicknesses even smaller, on the order of 1 nm), the situation may be different. We have calculated that insulating behavior occurs for very thin layers of VO$_2$ (one or two layers), and then a very peculiar zero-gap, point Fermi surface state emerges for 3 and 4 layers. 
Metallicity takes over for a thickness of 5 VO$_2$ layers and beyond.
This metal-insulator transition is not caused by, nor even accompanied by, any strong structural change. 
These IFs are non-polar, so this MIT differentiates very clearly from that of the LAO/STO system, its origin 
being completely different. Our analysis of the change in electronic structure with VO$_2$ layer thickness
indicates this is not an interfacial phenomenon 
but rather an effect of confinement of electronic states within the VO$_2$ layers coupled with specific 
orbital ordering on the V $d^1$ ions.

\section{Computational methods}

Our electronic structure calculations were  performed within density functional 
theory \cite{dft} using the all-electron, full potential code {\sc wien2k} \cite{wien}  
based on the augmented plane wave plus local orbital (APW+lo) basis set.\cite{sjo}
The exchange-correlation 
potential utilized to deal with possible strong correlation effects was the LSDA+U 
scheme \cite{sic1,sic2} including an on-site U and J (on-site Coulomb repulsion and exchange strengths) 
for the Ti and V $3d$ states.  The values U= 3.4 eV, J= 0.7 eV have been used for 
both Ti and V to deal properly with 
correlations in this multilayered structure; these values are comparable to (slightly smaller than)
what have been used for bulk VO$_2$\cite{vo2_peierls2,tomczak,haverkort} and other vanadates close to the itinerant limit.\cite{znvo_prl} Our calculations 
show that a larger U, of 4 eV, gives an incorrect insulating behavior of bulk VO$_2$ in 
rutile structure, hence overestimating electron-electron interactions.
Spin-orbit
coupling (SOC) effects are discussed in the last section.

\section{Results}

The multilayer system (TiO$_2$)$_m$/(VO$_2$)$_n$ (we will refer to it as m/n for simplicity), 
formed by $m$ TiO$_2$ monolayers and $n$ VO$_2$ 
monolayers grown along the rutile (001) direction, is calculated to undergo a MIT when 
$m$ is sufficiently large to confine VO$_2$ to two-dimensionality, that is, to prevent direct 
interactions between successive VO$_2$ slabs. Five layers of TiO$_2$ (or approximately 1.5 nm 
thickness) is more than sufficient, and that is what we use for our discussion. To follow 
this MIT we will analyze one by one different $n$ values from 1 to 5, from insulator for small 
$n$ to a metal for $n$ of 5 or greater. Our calculations include a full structural optimization 
($a$ and $c$ lattice constants as well as internal coordinates), starting from the rutile structure. 
Note that, because V is always $d^1$, there is one occupied band per V ion in the (super)cell.
For FM alignment, there will be $2n$ occupied bands in the majority structure for an $m/n$
superlattice.  
% For antiferromagnetic (AF) alignment, where double (spin) degeneracy persists, there will be only
% $n$ occupied bands for an $m/n$ superlattice (i.e. $n$ layer VO$_2$ slab). 

\begin{figure}[ht]
\begin{center}
\includegraphics[width=0.49\columnwidth,draft=false]{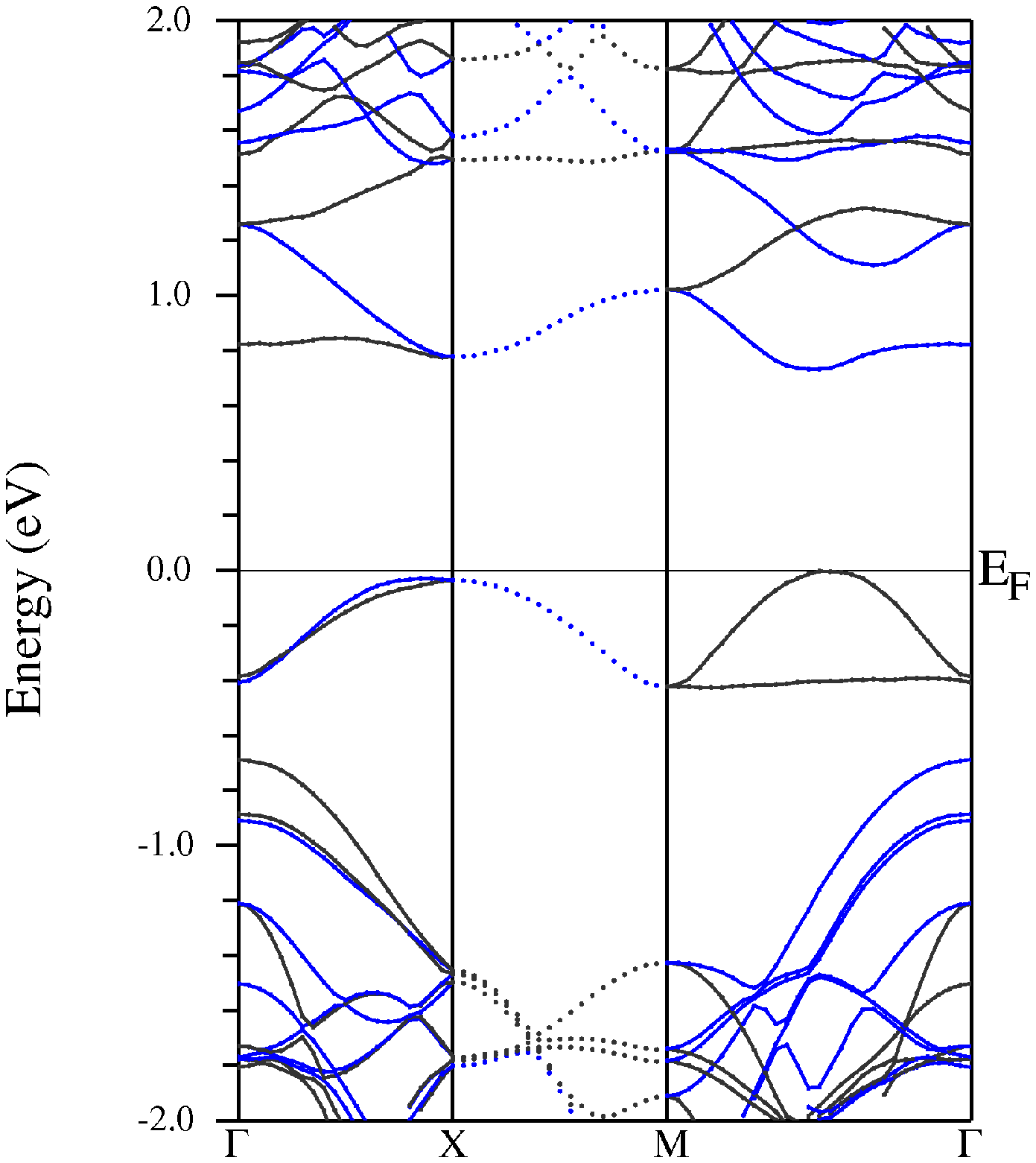}
\includegraphics[width=0.49\columnwidth,draft=false]{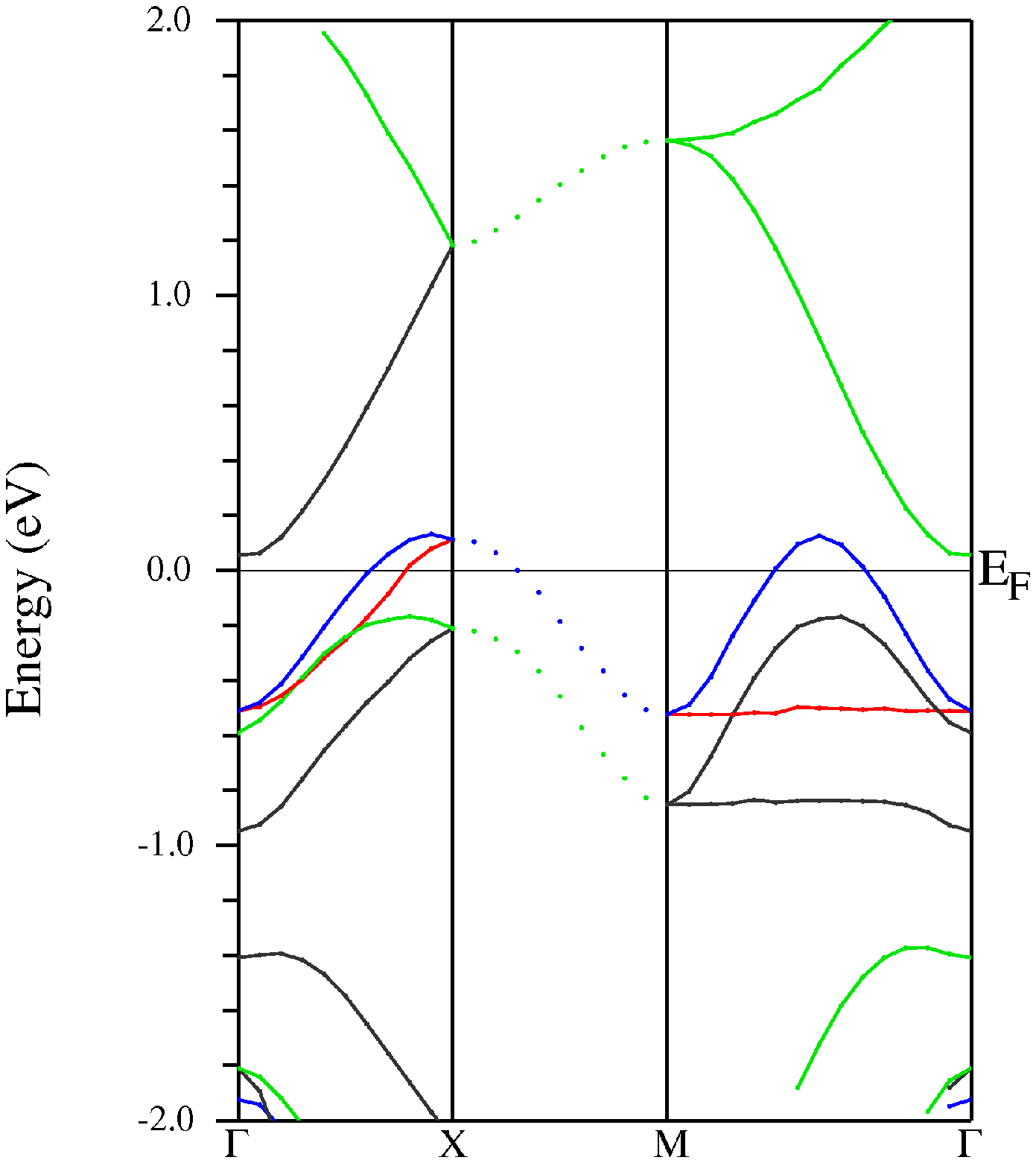}
\caption{(Color online) The left panel shows the band structure of 
the majority spin (TiO$_2$)$_5$/(VO$_2$)$_1$ 
multilayer along the symmetry lines in the $k_x, k_y$ plane. The right panel shows the majority
spin bulk bands 
in the VO$_2$ rutile structure when the Bloch wavefunction phase is constrained to be the same in 
neighboring cells along the $c$ direction (i.e. $k_z$=0). 
Observe the similarities in shapes between these two band structure of V $d$ bands in the -0.5 to 0.0 eV region 
of occupied states.}  \label{bs_5_1}
\end{center}
\end{figure}

To understand the structure of the multilayer, we need to consider that rutile structure with two V ions
per unit cell produces a corrugated interface between the two oxides. Hence, that leads to two structurally 
inequivalent V atoms at the interface, which have different coordinations with Ti ions across the interface.

\subsection{Single VO$_2$ cell slab}
To understand the evolution of the electronic structure with number of VO$_2$ layers, we focus primarily
on the ferromagnetically aligned cases.
Figure \ref{bs_5_1} shows the insulating band structure of the majority bands of the 5/1 
system for FM alignment of the moments. 
This shape of the occupied bands, which
is essentially that of the interfacial V atom band in the 5/3 system that we discussed 
in a previous paper,\cite{sD_prl} underpins the understanding of the change in electronic structure as $n$ increases. 
One band is flat, or nearly so. The other of the pair has a sinusoid shape, although the end segments are
not as quadratic as a true sinusoid. The two bands are degenerate at both $\Gamma$ and at M.
Although these bands have a simple shape and would be expected to have a simple representation in
terms of a tight-binding model, the two orientations of the VO$_2$ octahedra in the rutile structure,
and their lack of alignment of their natural axes with each other and with the cubic axes, preclude
a straightforward representation by a tight-binding fit. 

The gap of 0.8 eV present in this band structure can be attributed to its Mott insulating character;
each V ion has a single $3d$ orbital occupied.
The occupied band width is 0.4 eV; all other majority states lie above the gap.
For comparison, in the right panel of Fig. \ref{bs_5_1} we show the band structure of bulk VO$_2$ 
in rutile structure, calculations being run with the same value of U. The V $d$ 
bands have the same shape and symmetry in both systems, but the bandwidth is reduced when the system is 
confined within TiO$_2$ by a factor 2 for the occupied bands. The unoccupied bands in the multilayer 
system become complicated by the presence of Ti $3d$ bands in the same energy region.

\subsection{Bilayer VO$_2$ slab}
\begin{figure}[ht]
\begin{center}
\includegraphics[width=\columnwidth,draft=false]{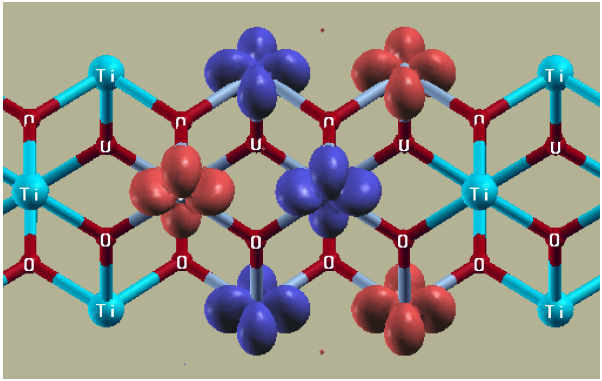}
\caption{(Color online) Spin density isosurface plot at 0.12 e/\AA$^3$ for the n= 2 nanostructure. 
The occupation at both V sites is the d$_{\parallel}$ orbital, with antiferromagnetic coupling
leading to the ``spin singlet'' state. 
Different colors represent different spin orientations.}\label{rho_5_2}
\end{center}
\end{figure}

Increasing the thickness to two layers of VO$_2$ (Fig. \ref{rho_5_2}), both FM and AF couplings along the rutile c-axis can be studied, with the possibility of forming some precursor of the V-V dimers that occur in the dimerized monoclinic insulating low temperature phase of VO$_2$.

The 5/2 system is found to be an insulator, whether the V spin alignment is FM or AF. 
The energetically favored state, by 9 meV/V, has the two V atoms along a V chain 
(confined within 5 layers of TiO$_2$) 
coupled antiferromagnetically.
As found in our previous report\cite{sD_prl} for 
the 5/3 system, the AF magnetic coupling leads to occupation of the $d_{\parallel}$ orbital, 
with one pair of lobes directed along the c-axis, and forming the 
$dd\sigma$ bonds along the rutile c-axis that can be seen in Fig. \ref{rho_5_2}. The plot 
presents the spin-density isosurface and allows to visualize the $d_{\parallel}$ orbital
from the V$^{4+}$ cation. This electronic structure is analogous to that of VO$_2$ at 
low temperatures in the monoclinic dimerized phase, when it transitions to an insulating 
phase from the high temperature metallic rutile structure, except there is no dimerization. 
In VO$_2$, the $\sigma$-bond along the chain and dimerization leads to a spin-singlet 
formation along the chain, resulting in a gap. In the case of multilayers of the thickness we
consider (and even larger), the effect is not cooperative and no temperature-induced 
metalization is observed, as mentioned earlier.

The FM configuration of the 5/2 system is also insulating, but with a smaller band gap than in
the AF case. To allow comparison of a systematic series of nanostructures, the band structure 
of the 5/2 system in a FM state along the rutile c-axis is included as the left panel of 
Fig. \ref{bs_5_n}. When comparing this 5/2 system to  5/1 and 5/3, it can be seen how the 
bands that cross at the semi-Dirac point in the 5/3 system are approaching this energy region 
already in the 5/2 system.  The FM configuration in this 5/2 case is not the lowest energy state, 
but it will be useful to see this reference band structure for understanding the origin of the 
semi-Dirac point Fermi surface that appears in the 5/3 and 5/4 ML systems, as shown in the middle
two panels of Fig. \ref{bs_5_n}. 
When comparing the 5/2 and 5/1 band structures, we see a new set of two V occupied bands, similar 
to those present in the 5/1 band structure, but with smaller bandwidth, 
and a corresponding reduction in the band gap of the full multilayer system can be observed. 
It is already clear that adding more layers of VO$_2$ will lead to a metallic state.

\subsection{Larger slabs, n= 3, 4, 5}

\begin{figure*}[ht]
\begin{center}
\includegraphics[width=0.52\columnwidth,draft=false]{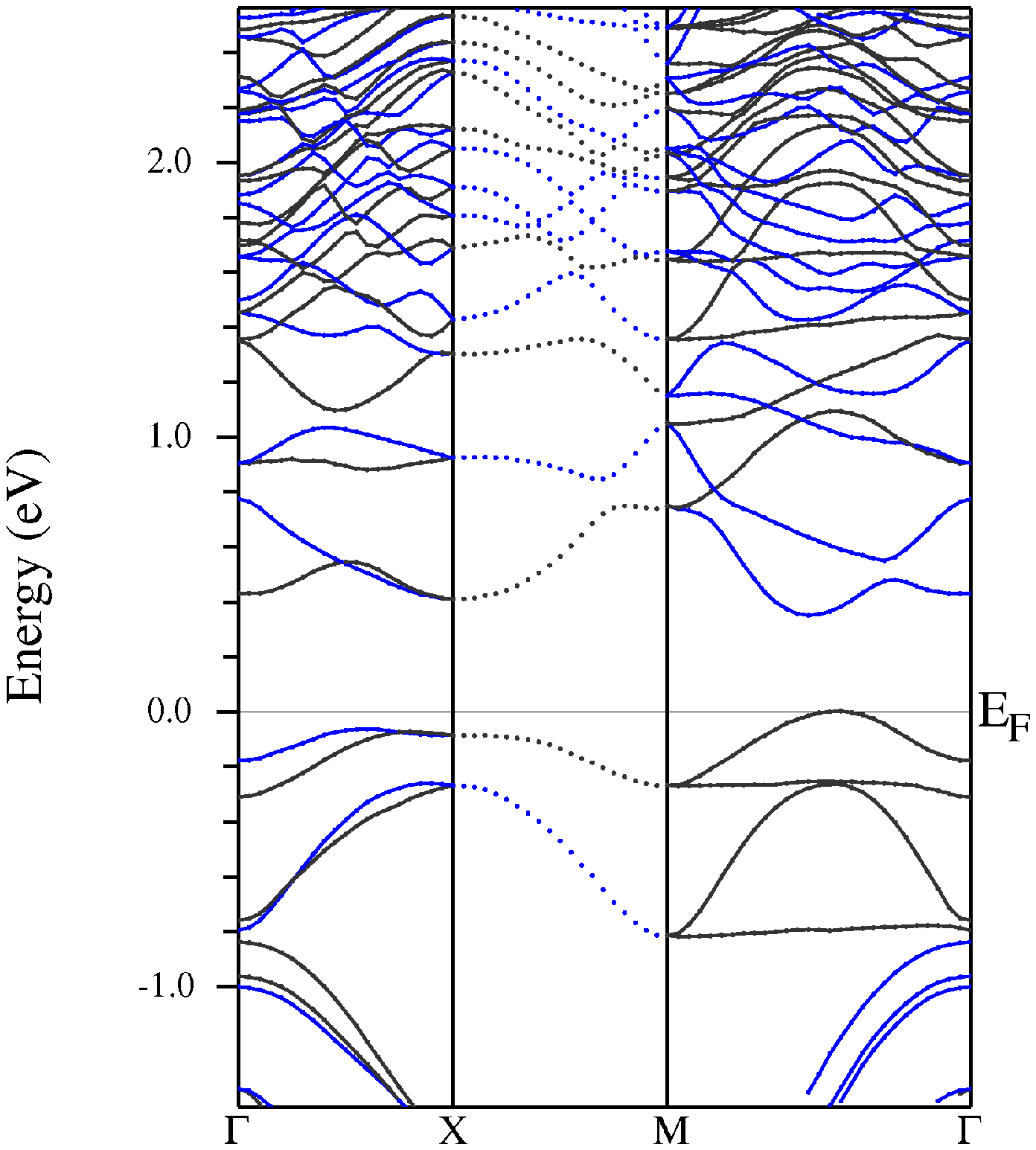}
\includegraphics[width=0.45\columnwidth,draft=false]{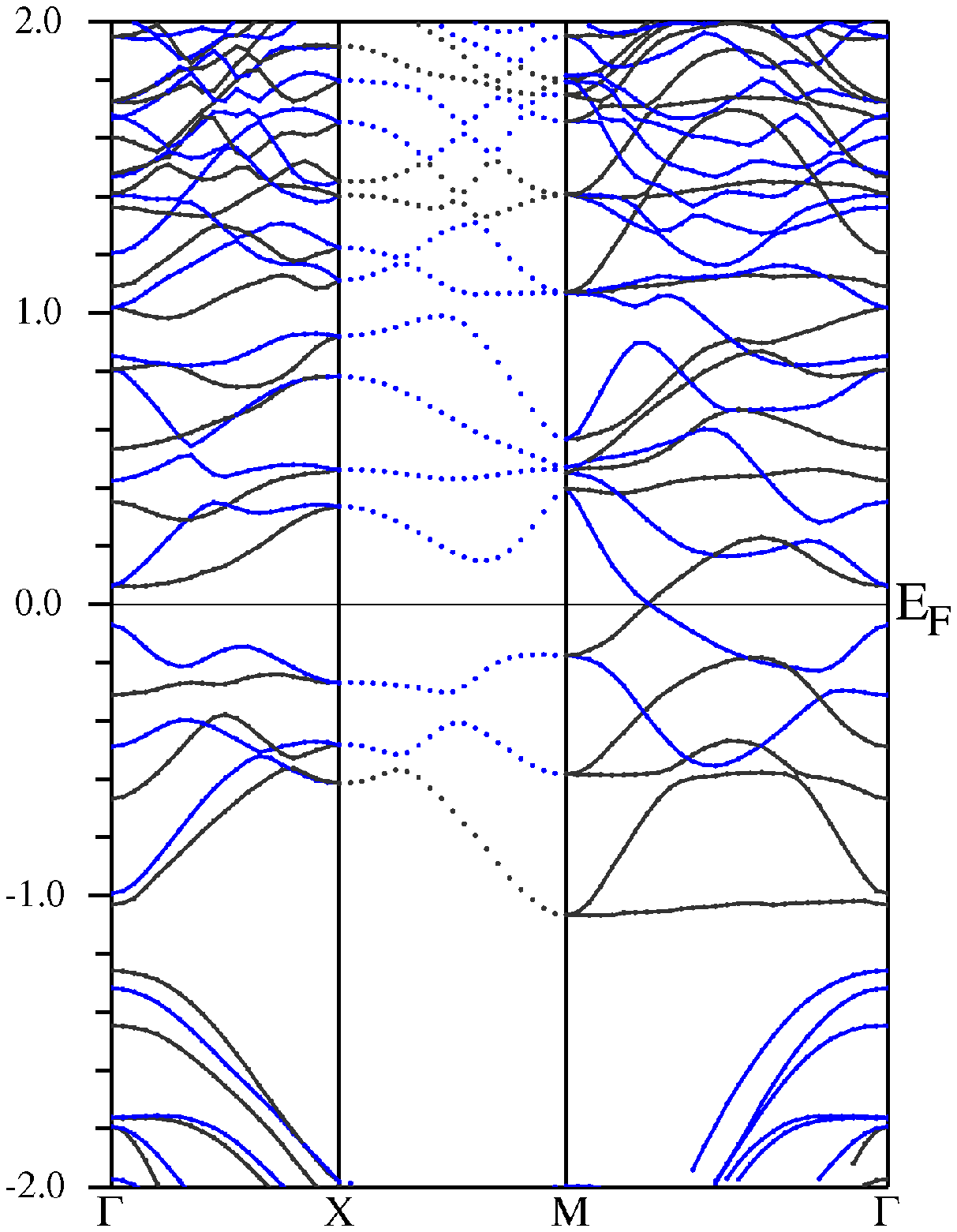}
\includegraphics[width=0.45\columnwidth,draft=false]{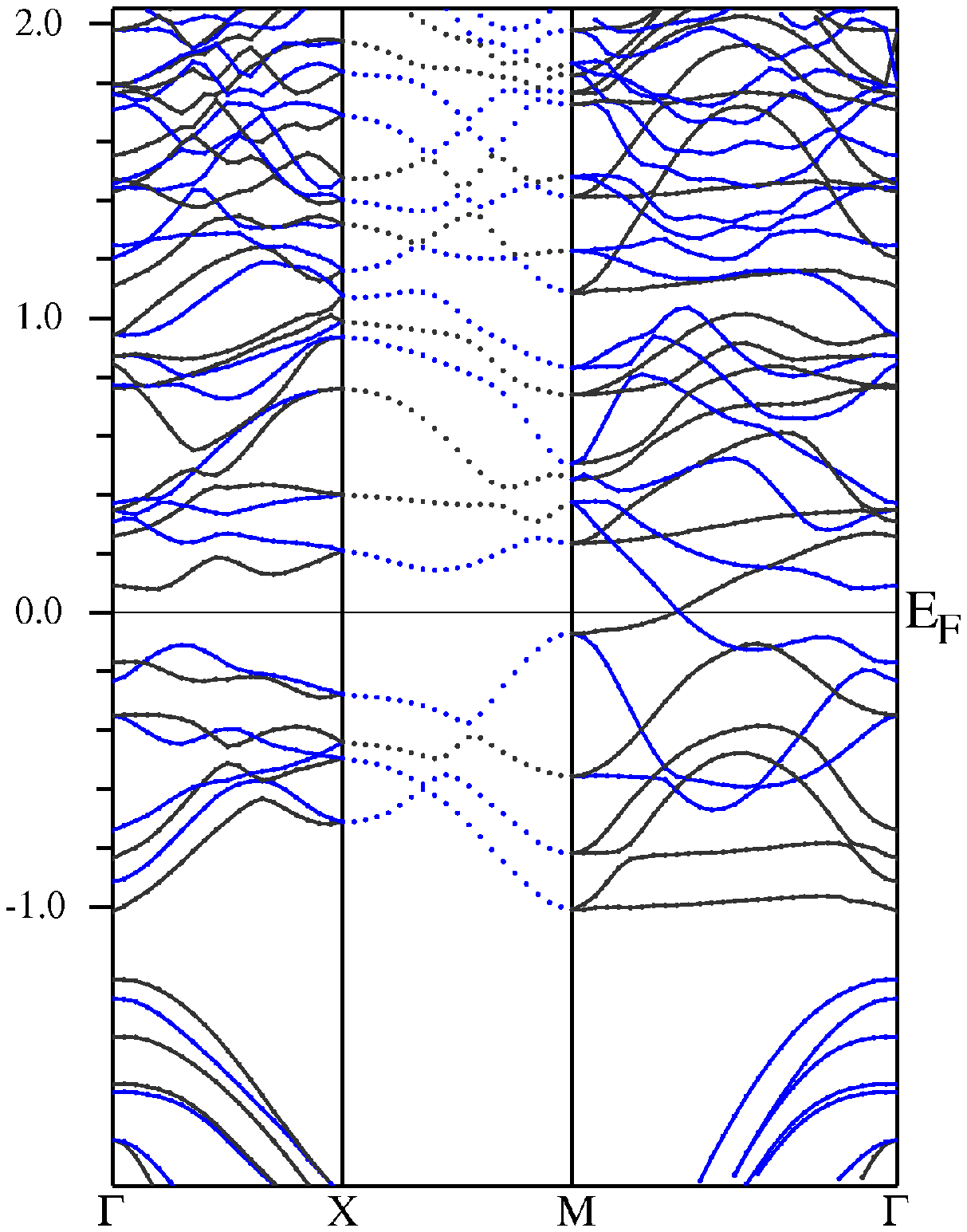}
\includegraphics[width=0.45\columnwidth,draft=false]{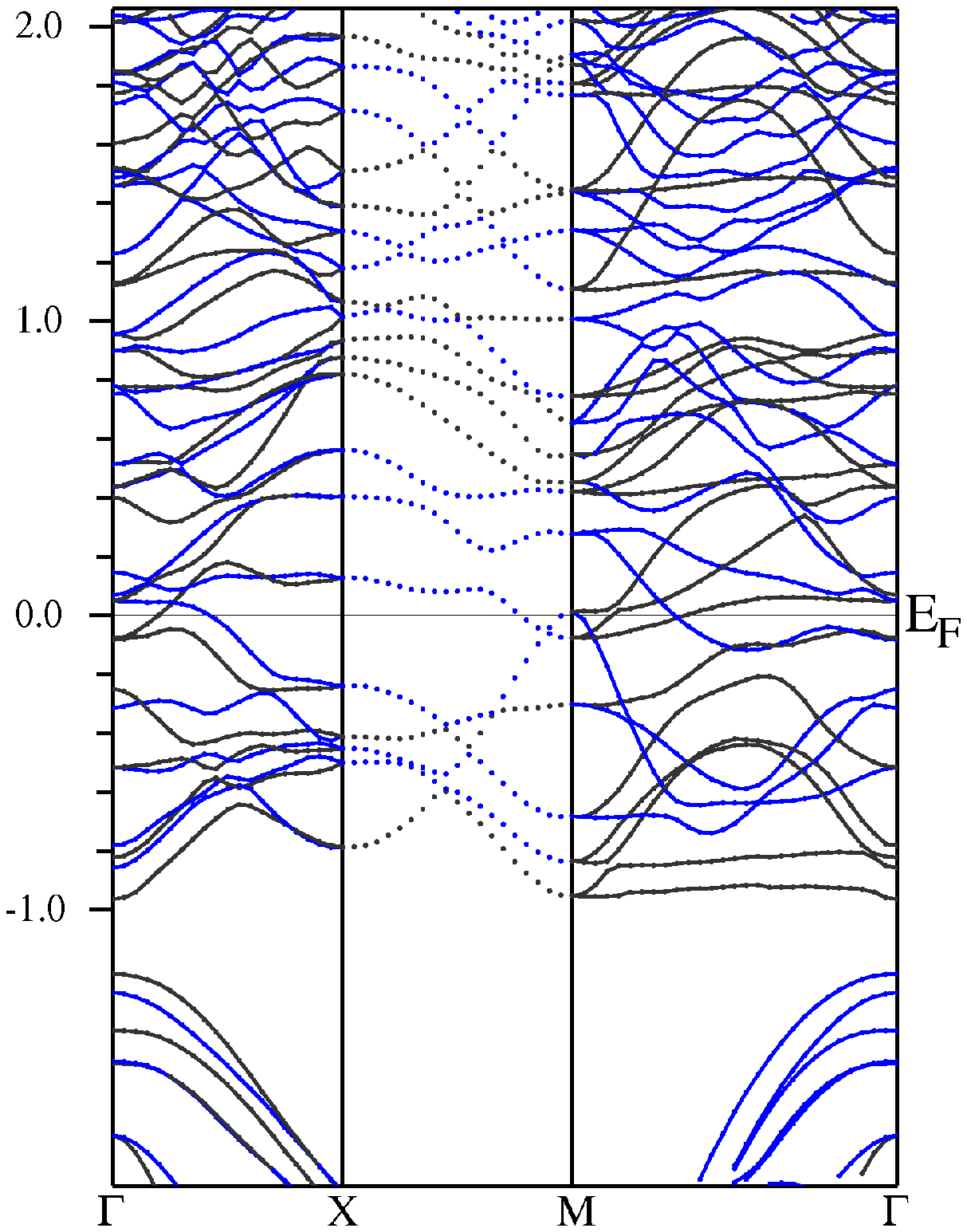}
\caption{(Color online) Band structure of the (TiO$_2$)$_5$/(VO$_2$)$_n$ multilayer 
system for n from 2 (left) to 5 (right). This shows how the transition to a metallic from 
an insulating state occurs through the appearance of a semi-Dirac point for n= 3 and 4. 
The band structure is represented on the k$_x$ and k$_y$ plane, enough for the 
two-dimensional symmetry of the system.}\label{bs_5_n}
\end{center}
\end{figure*}

To provide the sequence up to, at, and beyond the semi-Dirac cases, Figure \ref{bs_5_n} shows 
the band structures of the (TiO$_2$)$_5$/(VO$_2$)$_n$ system for values 
of $n$ from 2 to 5, from left to right. 
The alignment of energies in the various band structures presented in this figure was carried out using 
as a reference the O $1s$ core state of the central oxygen in the (TiO$_2$)$_5$ layer. Using that reference, 
the occupied oxygen bands coming from the TiO$_2$ layer are about the same position in energy, but the 
Fermi levels are not aligned. Nevertheless, the Fermi levels for $n$ = 3, 4, 5 are very nearly aligned,
as would be expected once a metallic band structure is established.

\begin{figure}[ht]
\begin{center}
\includegraphics[width=0.98\columnwidth,draft=false]{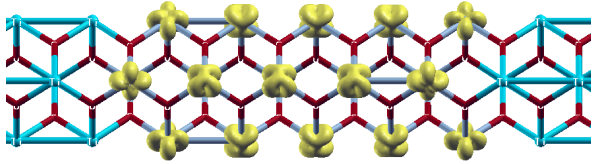}
\caption{(Color online) Spin density isosurface of (TiO$_2$)$_5$/(VO$_2$)$_5$ multilayer system at a value of 0.25 e/\AA$^3$. Observe that the non-interfacial V atoms have a different electronic structure, and a large orbital angular momentum is expected for them.}\label{rho_5_5}
\end{center}
\end{figure}

As presented and discussed earlier\cite{sD_prl} the 5/3 system displays a very peculiar electronic structure not seen previously,
consisting of a semi-Dirac dispersion around the point Fermi surface.  The dispersion is massless (linear)
along the (1,1) symmetry direction in k-space, and massive (quadratic) perpendicular to this direction at the semi-Dirac
point. Such an unusual dispersion will lead to 
unexpected and unique physical properties,\cite{swapno_sD} many of which remain to be studied. The origin of 
these peculiar Fermi points arises from a specific coupling of two bands related to the change in orbital ordering
from the first unit cell to deeper cells.  The change in orientation is the same in the 5/3 system as is 
pictured in the 5/5 system in Fig. \ref{rho_5_5}. The interface V orbital always has the orientation shown
in Fig. \ref{rho_5_2} for the 5/2 system. The orientation of the inner atoms is that of FM bulk VO$_2$.

There are two pairs of bands, each with its own identifiable shape, that can be used to 
follow the formation and then the disappearance of the 
semi-Dirac band structure.  The occupied pair in the 5/1 system was discussed above, and it appears twice
in the 5/2 system, with centers displaced by about 0.4 eV.
In the 5/3 system, a new pair of bands appears, which their character reveals to be a 
mixture of bands coming from V sites that are {\it not} at the interface. This pair of bands can be 
identified also in the 5/2 band structure, where they are the lowest pair of unoccupied states at the
M point.  These bands disperse in opposite
directions from the M point, and the upper member crosses a band that is dispersing downward from above.
This crossing gives rise to the semi-Dirac point.
Since the crossing bands do not involve the interface V sites, and does not survive for more than four unit cells
of VO$_2$, the semi-Dirac phenomenon qualifies 
as a quantum confinement effect
rather than an interfacial effect. 

The occupied V orbital away from the interface is largely (but certainly not purely)
a $d_{x'z'}$$\pm$ i $d_{y'z'}$ orbital
in a coordinate system that has $z'$ in the $a-b$ plane. By symmetry, it is rotated by 90$^{\circ}$ for the other
V ion in the cell. The distortion of the octahedron, and the two orientations of octahedra, in the rutile
structure, make this orbital difficult to quantify in terms of the cubic $3d$ orbitals, and also complicate the
tight-binding representation of what appear to be simply dispersing pairs of bands.
More comment on these orbitals are provided in Sec. \ref{soc} below.  The two types of orbital
occupation lead to the dispersion that 
finally produces the semi-Dirac point that pins the Fermi level.

\begin{figure}[ht]
\begin{center}
\includegraphics[width=9.2cm,draft=false]{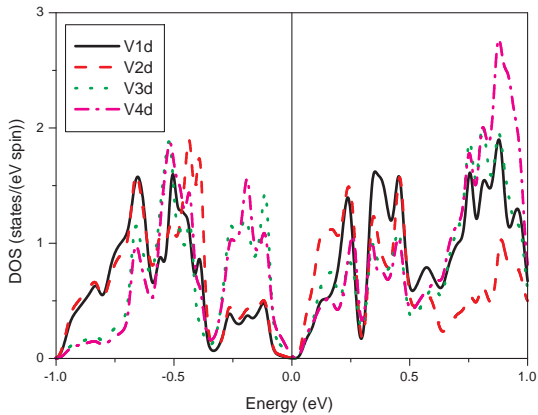}
\caption{(Color online) Density of states of the (TiO$_2$)$_5$/(VO$_2$)$_4$ multilayer system. This shows a zero-gap situation, with most of the spectral weight around the Fermi level coming from the inner V atoms, away from the interface. Naming convention along the rutile c-axis is V1-V2-V3-V4-V4-V3-V2-V1, V1 (V2) being the outermost (innermost) interfacial atom.}\label{dos_5_4}
\end{center}
\end{figure}

The 5/4 system is closely related to that of the 5/3 system. As in the 5/3 nanostructure, the FM configuration
is lower in energy than an antiferromagnetic alignment along the rutile c-axis, by 57 meV/V.  
This disfavoring
of AF ordering is consistent
with the experimental observation that long-range dimer formation along the rutile c axis only
occurs at VO$_2$ thicknesses of several nm,\cite{vo2_jap} enough to sustain a dimerization and structural cooperative
distortion of the rutile-based lattice.
Just as in the 5/3 system, this nanostructure has a half metallic zero-gap semi-Dirac structure.
Figure \ref{dos_5_4} shows the partial density of states (DOS) of
the four inequivalent V atoms, and the two central cells are different from those at the interface.   
For labeling the atom-projected DOS curves, the V atoms are named as
V1-V2-V3-V4-V4-V3-V2-V1 from interface to interface along the rutile $c$ axis. The two
innermost V atoms V3 and V4 contribute most to the DOS near the Fermi level, but very near
the Fermi level a
contribution from V2 (the more inner atom in the interfacial unit cell) becomes noticeable. The 
interfacial V1 atom has no contribution in either of the two bands that cross the Fermi level.
Again, the semi-Dirac point is a confinement effect.

Increasing the thickness of the VO$_2$ layer by adding a fifth layer to create the 5/5 multilayer, 
more bands appear at the Fermi level and the system reaches a metallic state, as shown in the right panel
of Fig. \ref{bs_5_n}.  Thus the insulator-metal transition has completed for the thickness of the 
VO$_2$ layer greater than 4 layers. The transition has proceeded through a very peculiar point-like 
Fermi surface stage that takes place only in the 5/3 and 5/4 systems.  In the 5/5 system a crossing of 
two semi-Dirac bands occurs near the Fermi level along the $\Gamma$-$M$ direction as in the 5/3 and 5/4
systems, but the system is metallic (normal) because several other bands also cross the Fermi level.

\subsection{Band line-up across the interface}
\begin{figure}[ht]
\begin{center}
\includegraphics[height=4.8cm,draft=false]{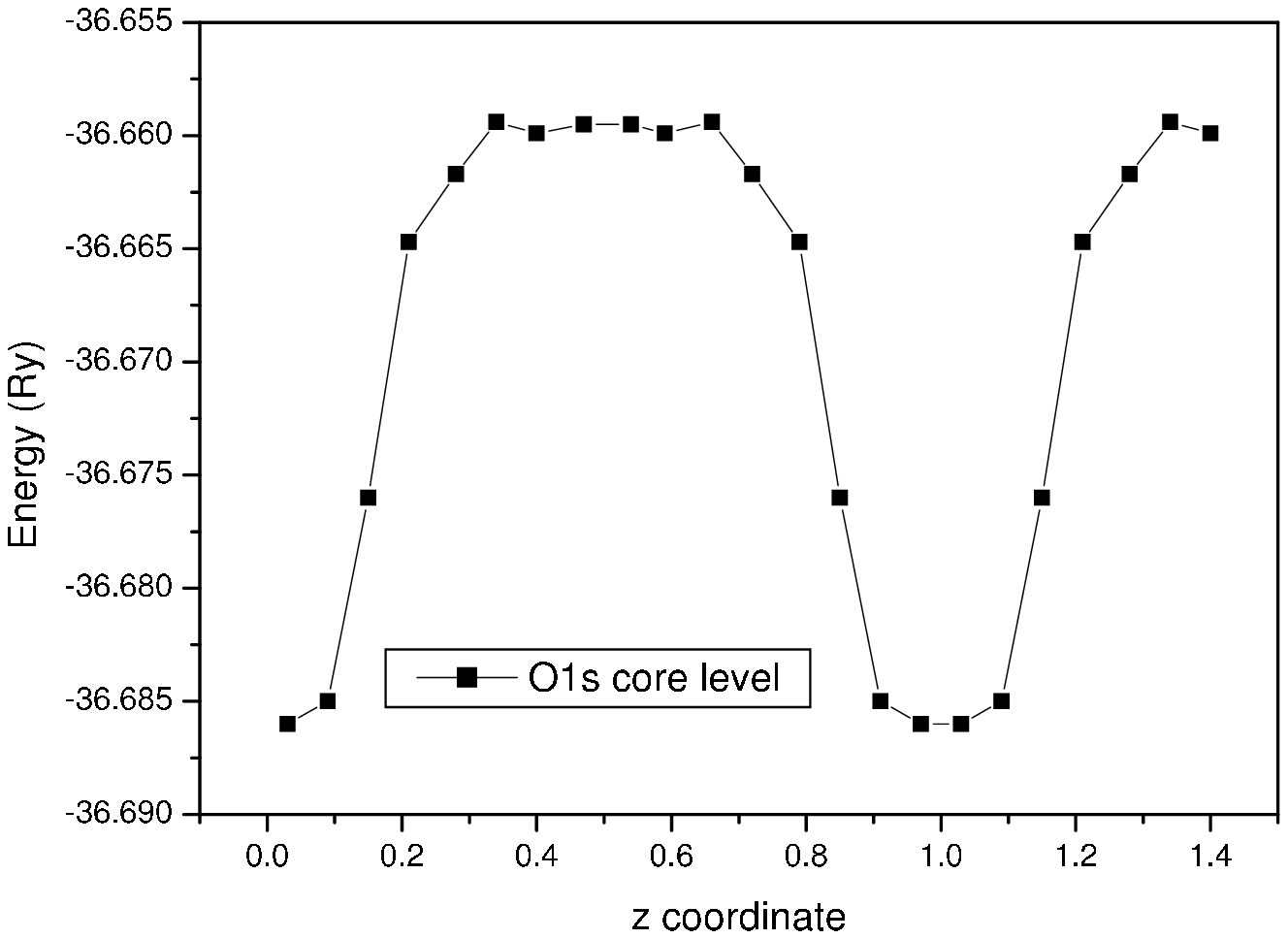}
\vspace{0.3cm}
\includegraphics[width=\columnwidth,draft=false]{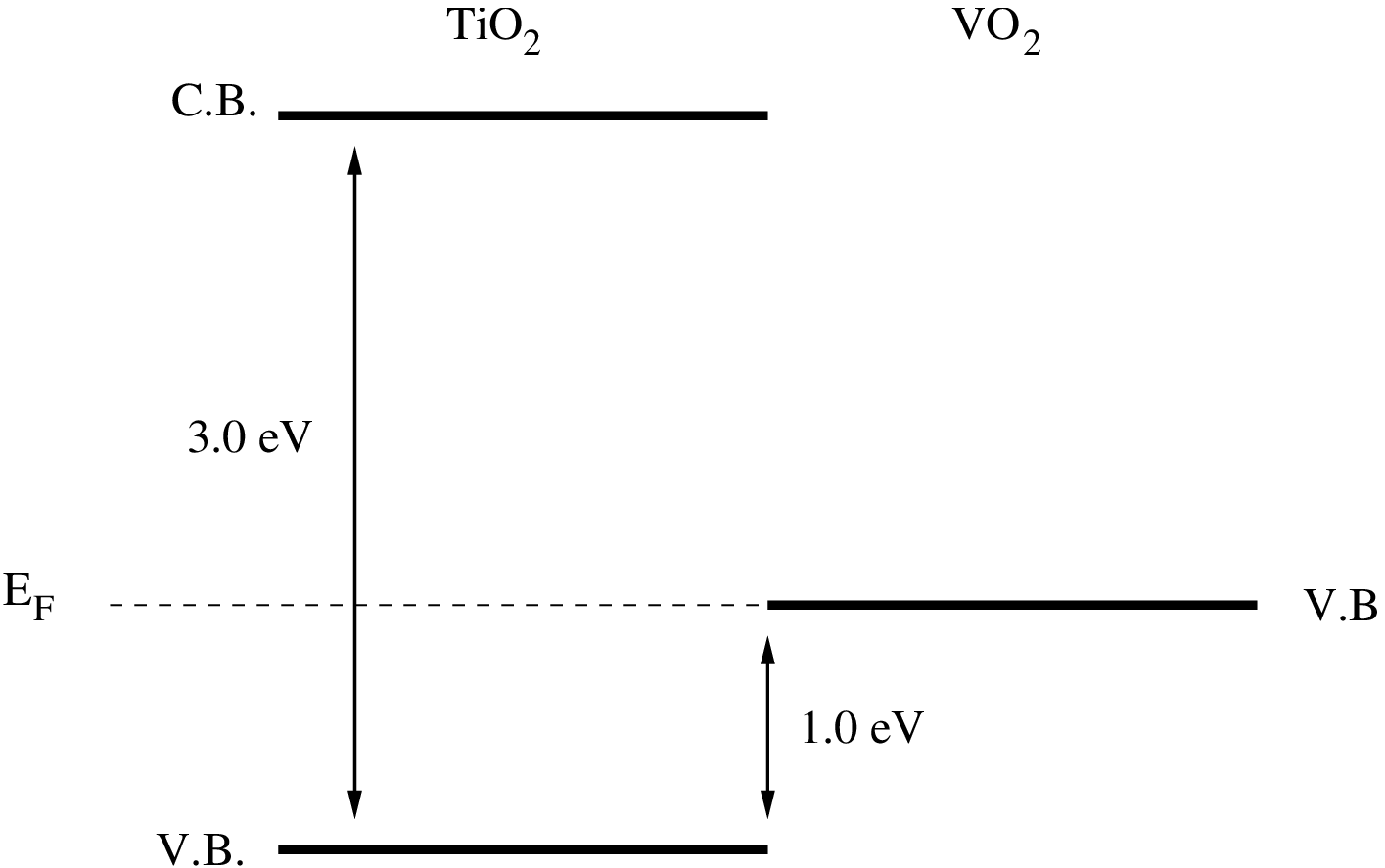}
\caption{Top panel: O $1s$ core level energies, plotted versus the $z$ position across the 
multilayer. The valley corresponds to the VO$_2$ layer and the hill is the TiO$_2$ layer.
Bottom panel: band alignments across the interface, showing that the Fermi level falls 1.0 eV above
the bottom of the 3.0 eV TiO$_2$ band gap.}\label{si_1}
\end{center}
\end{figure}

We use the 5/3 system to obtain the band line-up across the TiO$_2$/VO$_2$ interface, 
by referencing to the O $1s$ core levels and then using the bulk electronic structures. 
The top panel of Fig. \ref{si_1} shows that the O $1s$ core level position converges 
(to the bulk value) within the TiO$_2$ layer, but perhaps is not quite converged in the thinner VO$_2$ layer. 
Still, this amounts to only a minor correction of the band alignments for an isolated interface.  
The resulting band
line-up is shown in the lower panel of Fig. \ref{si_1}: the Fermi level of VO$_2$ lies 1.0 eV
above the bottom of the 3.0 eV gap of TiO$_2$. 

The behavior of the $1s$ level energies provides insight into the healing length of
the local potential away from the interface.  In Fig. \ref{si_1}, the center of the TiO$_2$
layer lies at $z$=0.5 and the center of the VO$_2$ layer at $z$=1.0.  The three central layers
(of five total) of TiO$_2$ have the same local potential, hence it is only the single TiO$_2$
unit cell at the interface that experiences the effect of the interface on the potential.
Only the cell at the interface is affected also in VO$_2$, however, the energy difference between
the two O ions in the cell is almost four times greater than on the TiO$_2$ side. This
result is consistent with a more polarizable $d^1$ ion compared to the $d^0$ ion, {\it i.e.}
more screening on the VO$_2$ side. 

\section{The semi-Dirac point}
The band structure of the 5/4 system (third panel of Fig. \ref{bs_5_n}) shows the appearance of 
pair of bands crossing just at the Fermi level, at only a single point along the plotted symmetry
lines; {\it i.e.} a potential semi-Dirac point.\cite{sD_prl,swapno_sD} It lies along
the $M-\Gamma$ direction, just as in the 5/3 case and in fact is similarly placed along the zone 
diagonal. This semi-Dirac-point low energy electronic structure therefore displays robustness: not only
against variations in atomic positions and reasonable changes in the Coulomb repulsion
U, but also against interfacial disorder\cite{sD_prl}and the more substantial change of adding 
one additional layer of VO$_2$.  
For four VO$_2$ layers, the V $3d$ bands become narrower and have correspondingly more 
states in the same energy
region, but the dispersion typical 
of the rutile metallic VO$_2$ system remains. The other sets of bands, which come from the 
more internal VO$_2$ layers, 
are very similar to the 5/3 case.  

\begin{figure}[ht]
\begin{center}
\includegraphics[width=\columnwidth,draft=false]{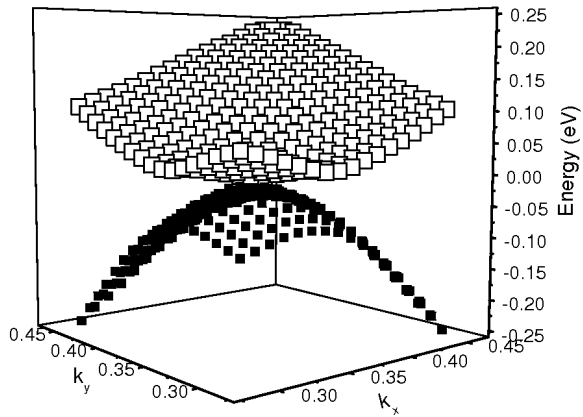}
\caption{ Band structure of the (TiO$_2$)$_5$/(VO$_2$)$_4$ multilayer system in a two-dimensional 
grid in the k$_x$-k$_y$ plane. The surfaces reveal a point Fermi surface, and also the semi-Dirac 
character of the dispersion. The Fermi level is set at zero energy.}\label{bs_3D_5_4}
\end{center}
\end{figure}

Since the band structure along symmetry lines does not necessarily determine what the 
dispersion is like (linear or
quadratic) perpendicular to the line, it is necessary to check explicitly. 
Figure \ref{bs_3D_5_4} shows a representation $\varepsilon$($k_x,k_y$) of the two bands crossing the Fermi 
level in the 5/4 system, from which the parabolic energy dispersion in the perpendicular
direction is evident.  
Hence the band crossing is indeed a semi-Dirac point, similar in all respects to the 5/3 system:  electrons 
at the Fermi level behave as relativistic massless fermions along the zone diagonal and as massive 
particles along the perpendicular direction. The behavior of the system of semi-Dirac fermions for 
the 5/3 system will apply here as well.\cite{sD_prl,swapno_sD}

\section{Spin-orbit coupling}\label{soc}

\begin{figure*}[ht]
\begin{center}
\includegraphics[width=0.98\columnwidth,draft=false]{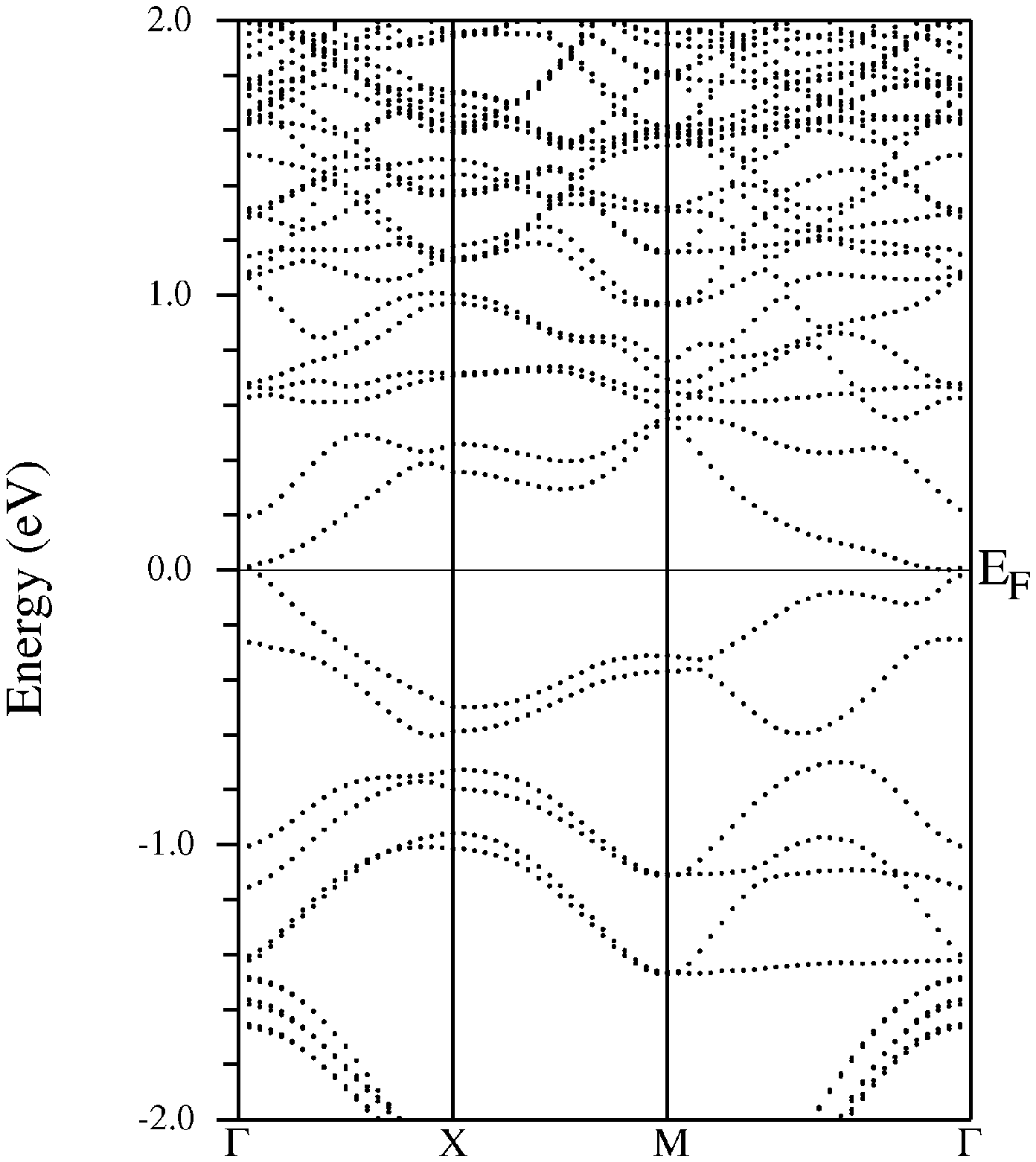}
\includegraphics[width=0.98\columnwidth,draft=false]{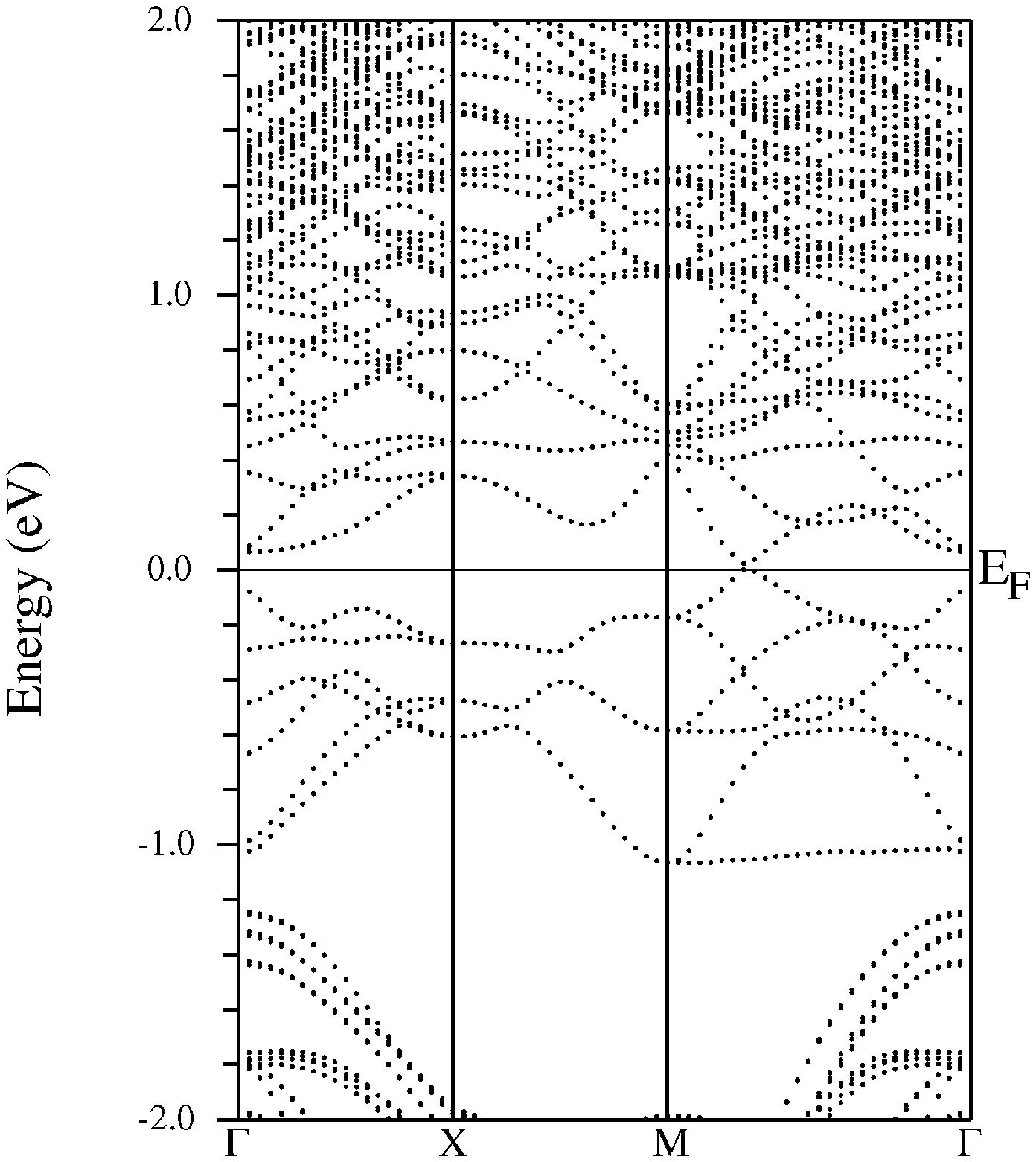}
\caption{(Color online) Band structure of the (TiO$_2$)$_5$/(VO$_2$)$_3$ multilayer system when spin-orbit coupling is introduced. This destroys the semi-Dirac point for the case of magnetization in the xy plane (left), but not for the ground state case of the magnetization along the rutile c axis (right), where the semi-Dirac point persists.}\label{bs_5_3_so}
\end{center}
\end{figure*}

One important question that needs to be addressed is how
spin-orbit coupling (SOC) affects the bands in the vicinity of the semi-Dirac point.
At the Dirac point of graphene, it is known that SOC leads to a very small gap opening, destroying the perfection of the Dirac point.\cite{graphene_soc} However, the gap is
so small that it probably has no measurable consequences. V atoms are somewhat heavier,
so SOC should have a larger effect.  There are in fact several interesting complications
in this system.

First of all, this three-cell layer of VO$_2$ is not only magnetic (unlike graphene),
but it is half-metallic.  The nearest minority bands that can be mixed by SOC into the
majority bands at the semi-Dirac point are 1 eV away.  Mavropoulos {\it et al.} \cite{mavropoulos}
have studied the effects of SOC in some 
intermetallic half metals, but without reporting on the size of any band gaps that might
have resulted.  They found that the amount of mixing (in terms of DOS) of minority into
the majority band gap could range from 0.5\% to a little more than 10\%, depending on the size
of the half metallic gap, the position of the Fermi level in the gap, etc.  As a result
of SOC, half metals are no longer strictly 100\% half metallic due to spin mixing.

Second, our system is strongly two dimensional, and even the underlying rutile structure
is anisotropic, especially in VO$_2$ where the spin coupling along the $c$-axis is
strongly tied to the structural transition and insulator-to-metal transition.\cite{vo2_09}  
However, unlike in graphene, there is a real underlying three-dimensional lattice in the
present case.  Thus
different effects may be observed depending on the direction of the V spin moments.
Thirdly, the two bands that cross at the semi-Dirac point have different weights on distinct
V ions, viz. mostly V2 ions for one, V3 ions for the other.  This will affect the magnitude
of the coupling, since SOC is a local (single atom) interaction.

The SOC operator in the Kohn-Sham Hamiltonian, 2$\times$2 in spin space, is
\begin{eqnarray}
H_{soc} = \xi \vec \ell \cdot \vec s = \xi[\ell_z s_z 
        + \frac{1}{2}(\ell_+ s_{-} + \ell_{-} s_{+})]
\end{eqnarray}
where $\ell_{\pm} = \ell_x \pm i \ell_y$ etc. 
We consider the two distinct 
cases of direction of the spins, and we are interested in the majority (spin ``up'') bands.
We focus on the
coupling only between the two bands ($a, b$) near the semi-Dirac point. In each case we quantize
$\vec \ell $ (hence $\vec s$) along the direction of the V spin moment. Then the matrix
element directly coupling the two bands becomes (suppressing the $k$ index)
\begin{eqnarray}
<a|H_{soc}|b> = \frac{\xi_V}{2} <a|\ell_z|b>.
\end{eqnarray}
The band-diagonal matrix elements, giving simple band shifts for the two bands, will 
not impact this gap opening question. The remaining coupling arises from second (and higher)
order terms involving the spin-down conduction band 1 eV and higher in energy, leading
to corrections that are at a maximum $\xi (\xi$/1 eV) $<\ell>^2$ which will be very
small, and we do not consider these corrections further (though they are included in
the band structure calculations).  From the above equation, band mixing will give a gap
$\sim$ $\xi_V$ $|<a|\ell_z|b>|$.

\noindent {Case 1. \it $\vec s$ along the $c$ axis.} 
This is the calculated ground state, with energy about 10 meV/V lower than for 
polarization in the plane.  In this case, the calculated band structure shows that
the semi-Dirac point at the Fermi level persists (right panel of Fig. \ref{bs_5_3_so}). 
The calculated value of orbital moment is 0.04 $\mu_B$ for each V ion, a fairly typical
moment for a $3d$ atom in a magnetic material.  The fact that the band crossing still
occurs indicates that SOC does not interfere with the difference of symmetries of the
two bands along the $k_x=k_y$ line that allows them to cross (without SOC); the matrix
element vanishes due to the differing symmetry of the bands. 

\noindent {Case 2. \it $\vec s$ along the $a$ axis.}
In this case, a substantial gap $\sim 0.2$ eV is observed 
(left panel of Fig. \ref{bs_5_3_so}) to open
up at the former location of the semi-Dirac point. 
The V orbital moments that result, which are six separate values since the V ions
all become inequivalent, are as high as 0.2 $\mu_B$ inside the muffin-tin sphere.
The largest value occurs for the $d_{xz}$ $\pm$ $i d_{yz}$ orbital whose local
$z$ axis is parallel to $\vec s$ and is ready-made to respond with an orbital moment
along the $z$ axis. 
The change in symmetry of the system when the moment
lies in the plane results in band mixing, {\it i.e.} the matrix element no longer
vanishes due to symmetry.  The unexpectedly large SOC-induced gap is connected with the
large orbital moment.  

Another observation from Fig. \ref{bs_5_3_so} can be noted.  In the left panel,
after the gap has opened due to SOC, the upper two occupied V $3d$ bands lie in energy
roughly where they were without SOC (right panel).  The two pairs of V $3d$ bands at
lower energies, however, have moved lower in energy by 0.4 eV, an unusually large shift for
an effect of SOC.  We attribute this shift to the large orbital moments that are 
induced when the spin lies in the plane.

That the $t_{2g}$ subshell provides a representation for L=1 (not L=2) orbital moments
has long been known.\cite{epr,stevens,enough,lacroix,eschrig} Fairly recently several 
cases have come to light in which orbital moments can
actually be {\it quenched} within the $t_{2g}$ shell\cite{khaliulin} by
structure-induced or spontaneous symmetry lowering, or an orbital moment can
compensate a spin moment and prevent orbital ordering.\cite{kwanwoo} Other more complicated 
orders may appear in Sr$_2$VO$_4$, where the effects of spin-orbit may lead to a hidden 
octupolar order with vanishing expectation values for both L and S.\cite{sr2vo4}
In the V$^{4+}$ $t_{2g}$ system studied in this paper, an unexpectedly large orbital moment
is induced by the spin moment for the $d_{xz}$ $\pm$ $i d_{yz}$ orbital.

\section{Conclusions}

In this paper we have presented electronic structure calculations on the multilayer system 
(TiO$_2$)$_m$/(VO$_2$)$_n$, to study the effects of  nanostructuring a correlated material like 
VO$_2$ that in the bulk undergoes a metal-insulator transition, sandwiched by a band insulator 
like TiO$_2$. The system provides an example of
 a $d^1$ - $d^0$ interface with no polar
discontinuity, so many of the questions that exist for other oxide heterostructures are moot here.
By increasing the VO$_2$ layer thickness, the multilayer shows a metal-insulator transition. 
For n $\geq$ 5 layers, the system becomes metallic, being insulator for n= 1 and 2. The system 
is, experimentally, susceptible to the recovery of the bulk properties typical of VO$_2$ for n $\sim$ 
15 (specifically, the temperature-induced structural transition). We have studied values of n 
significantly smaller, where the effects of quantum confinement can be important.
The metal-insulator transition in the multilayer occurs through a semi-Dirac point phase for 
n= 3 and 4, in which the Fermi surface is a sngle point, around which the electrons behave as massless along 
the zone diagonal in k-space and as massive fermions along the perpendicular direction. 
In previous papers, we analyzed the semi-Dirac state for the n= 3 case, finding it to 
be robust against interfacial 
disorder and a reasonable range of U values. In this paper, we observe that it persists also 
for n = 4, hence showing some (limited) robustness to the VO$_2$ thicknesses. 
We point out that spin-orbit coupling does 
not destroy this semi-Dirac point, where the spin lies along the rutile $c$ axis. 
The ground state has small orbital angular moments, 
hence the changes in the band structure introduced by 
spin-orbit coupling are very small and the semi-Dirac survives spin-orbit coupling.

\section{Acknowledgments}

We acknowledge useful interactions with S. Banerjee and R. R. P. Singh during the course
of this work. This project was supported by DOE grant DE-FG02-04ER46111 and through interactions with
the Predictive Capability for Strongly Correlated Systems team of the Computational
Materials Science Network and a collaboration supported by a Bavaria-California Technology
grant.  V.P. acknowledges financial support from Xunta de Galicia (Human Resources Program).

\end{document}